\documentstyle[times,psfig,mathrsfs]{mn}

\newif\ifAMStwofonts
\AMStwofontstrue

\ifoldfss
  \ifCUPmtlplainloaded \else
    \NewTextAlphabet{textbfit} {cmbxti10} {}
    \NewTextAlphabet{textbfss} {cmssbx10} {}
    \NewMathAlphabet{mathbfit} {cmbxti10} {} 
    \NewMathAlphabet{mathbfss} {cmssbx10} {} 
  \fi
  \ifAMStwofonts
    \ifCUPmtlplainloaded \else
      \NewSymbolFont{upmath} {eurm10}
      \NewSymbolFont{AMSa} {msam10}
      \NewMathSymbol{\upi}     {0}{upmath}{19}
      \NewMathSymbol{\umu}     {0}{upmath}{16}
      \NewMathSymbol{\upartial}{0}{upmath}{40}
      \NewMathSymbol{\leqslant}{3}{AMSa}{36}
      \NewMathSymbol{\geqslant}{3}{AMSa}{3E}

       \let\le=\leqslant
       \let\ge=\geqslant
    \fi
  \fi
\fi 

\ifnfssone
  \newmathalphabet{\mathit}
  \addtoversion{normal}{\mathit}{cmr}{m}{it}
  \addtoversion{bold}{\mathit}{cmr}{bx}{it}
  \newmathalphabet{\mathbfit} 
  \addtoversion{normal}{\mathbfit}{cmr}{bx}{it}
  \addtoversion{bold}{\mathbfit}{cmr}{bx}{it}
  \newmathalphabet{\mathbfss} 
  \addtoversion{normal}{\mathbfss}{cmss}{bx}{n}
  \addtoversion{bold}{\mathbfss}{cmss}{bx}{n}
  \ifAMStwofonts
    \ifCUPmtlplainloaded \else
      \UseAMStwoboldmath
      \makeatletter
      \new@mathgroup\upmath@group
      \define@mathgroup\mv@normal\upmath@group{eur}{m}{n}
      \define@mathgroup\mv@bold\upmath@group{eur}{b}{n}
      \edef\UPM{\hexnumber\upmath@group}
      \new@mathgroup\amsa@group
      \define@mathgroup\mv@normal\amsa@group{msa}{m}{n}
      \define@mathgroup\mv@bold\amsa@group{msa}{m}{n}
      \edef\AMSa{\hexnumber\amsa@group}
      \makeatother
      \mathchardef\upi="0\UPM19
      \mathchardef\umu="0\UPM16
      \mathchardef\upartial="0\UPM40
      \mathchardef\leqslant="3\AMSa36
      \mathchardef\geqslant="3\AMSa3E

       \let\le=\leqslant
       \let\ge=\geqslant
    \fi
  \fi
\fi 

\ifnfsstwo
  \DeclareMathAlphabet{\mathbfit}{OT1}{cmr}{bx}{it}
  \SetMathAlphabet\mathbfit{bold}{OT1}{cmr}{bx}{it}
  \DeclareMathAlphabet{\mathbfss}{OT1}{cmss}{bx}{n}
  \SetMathAlphabet\mathbfss{bold}{OT1}{cmss}{bx}{n}
  \ifAMStwofonts
    \ifCUPmtlplainloaded \else
      \DeclareSymbolFont{UPM}{U}{eur}{m}{n}
      \SetSymbolFont{UPM}{bold}{U}{eur}{b}{n}
      \DeclareSymbolFont{AMSa}{U}{msa}{m}{n}
      \DeclareMathSymbol{\upi}{0}{UPM}{"19}
      \DeclareMathSymbol{\umu}{0}{UPM}{"16}
      \DeclareMathSymbol{\upartial}{0}{UPM}{"40}
      \DeclareMathSymbol{\leqslant}{3}{AMSa}{"36}
      \DeclareMathSymbol{\geqslant}{3}{AMSa}{"3E}

       \let\le=\leqslant
       \let\ge=\geqslant
    \fi
  \fi
\fi 

\ifCUPmtlplainloaded \else
  \ifAMStwofonts \else 
    \def\upi{\pi}
    \def\umu{\mu}
    \def\upartial{\partial}
  \fi
\fi

   \title[The chemical evolution of $\omega$ Cen]{The chemical evolution of 
          Omega Centauri's progenitor system}
   \author[D. Romano et al.]{Donatella Romano,$^{1}$\thanks{E-mail: 
           donatella.romano@oabo.inaf.it} Francesca Matteucci,$^{2}$ Monica 
           Tosi,$^{1}$ Elena Pancino,$^{1}$ \cr Michele Bellazzini,$^{1}$ 
	   Francesco R. Ferraro,$^{3}$ Marco Limongi$^{4}$ and Antonio 
	   Sollima$^{3}$\\
           $^{1}$INAF\,--\,Osservatorio Astronomico di Bologna,
                 Via Ranzani 1, I-40127 Bologna, Italy\\
           $^{2}$Dipartimento di Astronomia, Universit\`a di Trieste,
                 Via Tiepolo 11, I-34131 Trieste, Italy\\
           $^{3}$Dipartimento di Astronomia, Universit\`a di Bologna,
                 Via Ranzani 1, I-40127 Bologna, Italy\\
           $^{4}$INAF\,--\,Osservatorio Astronomico di Roma,
                 Via Frascati 33, I-00040 Monteporzio Catone, Italy}
   
\begin{document}

     \date{Accepted 2006 December 21. Received 2006 December 20; in original 
           form 2006 November 27}

     \pagerange{\pageref{firstpage}--\pageref{lastpage}} \pubyear{2006}

     \maketitle

     \label{firstpage}


   \begin{abstract}
   Chemical evolution models are presented for the anomalous globular cluster 
   $\omega$\,Centauri. After demonstrating that the chemical features of 
   $\omega$\,Cen can not be reproduced in the framework of the closed-box 
   self-enrichment scenario, we discuss a model in which this cluster is the 
   remnant of a dwarf spheroidal galaxy evolved in isolation and then 
   swallowed by the Milky Way. Both infall of primordial matter and 
   metal-enriched gas outflows have to be considered in order to reproduce the 
   stellar metallicity distribution function, the age-metallicity relation and 
   several abundance ratios. Yet, as long as an ordinary stellar mass function 
   and standard stellar yields are assumed, we fail by far to get the enormous 
   helium enhancement required to explain the blue main sequence (and, 
   perhaps, the extreme horizontal branch) stellar data. Rotating models of 
   massive stars producing stellar winds with large helium excesses at low 
   metallicities have been put forward as promising candidates to solve the 
   `helium enigma' of $\omega$\,Cen (Maeder \& Meynet, 2006, A\&A, 448, L37). 
   However, we show that for any reasonable choice of the initial mass 
   function the helium-to-metal enrichment \emph{of the integrated stellar 
   population} is unavoidably much lower than 70 and conclude that the issue 
   of the helium enhancement in $\omega$\,Cen still waits for a satisfactory 
   explanation. We briefly speculate upon possible solutions.
   \end{abstract}



   \section{Introduction}
   \label{sec:int}

   The globular cluster (GC) $\omega$\,Cen (NGC\,5139) is a unique window into 
   astrophysics (see Smith 2004, for a recent review). The estimated mass, 
   between (2.5~$\pm$ 0.3)~$\times$~$10^{6}$~M$_\odot$ (van de Ven et al. 
   2006) and 5.1~$\times$~$10^{6}$~M$_\odot$ (Meylan et al. 1995) -- or even 
   7.1~$\times$~$10^{6}$~M$_\odot$ (Richer et al. 1991), makes it the most 
   massive GC of the Milky Way. Its total mass better compares with that of a 
   small dwarf spheroidal galaxy such as Sculptor 
   (${\mathscr{M}}_{\textrm{\scriptsize{Sculptor}}}$~$\simeq$ 
   6.4~$\times$~$10^{6}$~M$_\odot$; Mateo 1998) rather than with that of a 
   typical Galactic GC 
   (${\mathscr{M}}_{\textrm{\scriptsize{GGC}}}$~$\sim$~$10^{5}$~M$_\odot$; 
   Harris 1996). The large degree of chemical self-enrichment observed in 
   $\omega$\,Cen giant and subgiant members also sets it apart from all the 
   other Galactic globulars.

   Indeed, $\omega$\,Cen is the only known GGC to exhibit a large degree of 
   chemical self-enrichment in all the elements studied (Freeman \& Rodgers 
   1975; Cohen 1981; Mallia \& Pagel 1981; Gratton 1982; Norris \& Da Costa 
   1995; Smith, Cunha \& Lambert 1995; Smith et al. 2000; Pancino et al. 
   2002). While star-to-star variations in the abundances of light elements 
   such as carbon, nitrogen, oxygen, sodium, magnesium and aluminum could be 
   due to proton-capture fusion reactions occurring during quiescent hydrogen 
   and helium burning within the giants themselves, variations in the 
   abundances of heavier species -- such as iron-peak and neutron-capture 
   elements -- should be immune to such processes and, therefore, likely 
   reflect patterns imprinted on the observed stars by previous stellar 
   generations (Lloyd Evans 1983; Cohen \& Bell 1986; Norris \& Da Costa 1995; 
   see also Gratton, Sneden \& Carretta 2004, for a comprehensive review on 
   abundance variations within GCs).

   The first indication that $\omega$\,Cen was chemically inhomogeneous came 
   from the large color width of the red giant branch (RGB) established in an 
   earlier photometric work by Woolley (1966). In the last decade, both 
   extensive spectroscopic surveys and wide-field photometric studies -- 
   mostly of giant $\omega$\,Cen members -- have definitely confirmed the 
   existence of several (up to five) discrete stellar populations covering a 
   large range in metallicity (Norris, Freeman \& Mighell 1996; Suntzeff \& 
   Kraft 1996; Lee et al. 1999; Hilker \& Richtler 2000; Hughes \& Wallerstein 
   2000; Pancino et al. 2000; Rey et al. 2004; Sollima et al. 2005a). The 
   metallicity distributions of (sub)giant stars and main sequence turnoff 
   (MSTO) stars look pretty much the same (Stanford et al. 2006). From their 
   spectroscopic and photometric MSTO data, Stanford et al. (2006) find that 
   the formation of $\omega$\,Cen took place most likely over 2--4 Gyr; both a 
   null age range and age ranges higher than 6~Gyr are deemed unlikely (cf. 
   previous works by Norris \& Da Costa 1995; Hilker \& Richtler 2000; Hughes 
   \& Wallerstein 2000; Smith et al. 2000; Pancino et al. 2002; Rey et al. 
   2004; Sollima et al. 2005a; see also Kayser et al. 2006). It has also been 
   found that the faint main sequence of $\omega$\,Cen splits into at least 
   two distinct branches (Anderson 1997; Bedin et al. 2004; Sollima et al. 
   2006b). The blue main sequence (bMS) contains $\sim$25 per cent of the 
   stars and is 0.3~$\pm$ 0.2~dex more metal-rich than the red one (rMS; 
   Piotto et al. 2005). These observations are most likely explained by an 
   anomalously high helium abundance of bMS stars of $Y \ge$~0.38 ($\Delta 
   Y/\Delta Z \ge$~70; Norris 2004; Piotto et al. 2005). Such an intriguing 
   possibility has been quantitatively investigated in the framework of 
   somewhat idealized scenarios for the formation and evolution of 
   $\omega$\,Cen (Bekki \& Norris 2006).

   It has been shown (Norris et al. 1997) that the metal-rich component of the 
   cluster, which is more centrally concentrated, has a smaller line-of-sight 
   velocity dispersion and a lower systemic rotation about the cluster's minor 
   axis than the metal-poor one. A more recent analysis by Sollima et al. 
   (2005b) confirms the trend of decreasing velocity dispersion with 
   increasing metal abundance in the metallicity range 
   $-$2.0~$<$~[Fe/H]~$<$~$-$1.0, but shows also that in the extreme metal-rich 
   extension of the stellar population, now better sampled thanks to new 
   generation instrumentation (Pancino et al. 2000), this decreasing trend is 
   reversed (see fig.~9b of Sollima et al. 2005b). Asymmetries in the 
   distribution and velocity of the stars could testify past accretion events 
   within $\omega$\,Cen (Ferraro, Bellazzini \& Pancino 2002; Pancino et al. 
   2003), but evidence for these is not definitive yet (Platais et al. 2003). 
   To further complicate the overall picture, the orbit of $\omega$\,Cen is 
   found to be strongly retrograde, almost coplanar with the Milky Way disc, 
   and to have small apogalacticon, unlike any known Galactic GC (e.g. 
   Majewski et al. 2000). At variance with most globulars, the relaxation time 
   for $\omega$\,Cen is very long, up to a few times 10$^9$~years in the core 
   and a few times 10$^{10}$~years at half-mass radius (Meylan et al. 1995; 
   van de Ven et al. 2006). Indeed, recent results suggest that the cluster is 
   not yet relaxed \emph{even in the central regions} (Ferraro et al. 2006). 
   This cluster could then mantain for a fairly long time the imprinting of 
   its initial conditions, thus allowing one to use the currently observed 
   distribution of stars of different populations in order to trace back the 
   cluster formation and evolution (e.g. Merritt, Meylan \& Mayor 1997).

   The large mass, spread in element abundances, flattened shape and rotation 
   all come close to the picture where $\omega$\,Cen is the surviving remnant 
   of a larger system. The chemical and kinematical segregations detected in 
   $\omega$\,Cen add even more relevance to the picture of a dwarf galaxy 
   progenitor being subject to accretion events, since such gradients are 
   present in nearly all of the Local Group dwarf spheroidals (e.g. Harbeck et 
   al. 2001; Tolstoy et al. 2004; Koch et al. 2006, to name a few). Several 
   authors have speculated on the possibility that $\omega$\,Cen is the naked 
   nucleus of a dwarf satellite galaxy captured into a retrograde Galactic 
   orbit many billion years ago (Dinescu, Girard \& van Altena 1999; Majewski 
   et al. 2000; Smith et al. 2000; Gnedin et al. 2002; Bekki \& Norris 2006), 
   following more general ideas on an accreted origin for GCs in our own as 
   well as external galaxies (Zinnecker et al. 1988; Freeman 1993). 
   Self-consistent dynamical models as well as N-body hydrodynamical 
   simulations can be found in the literature which succeed to reproduce the 
   main features of the cluster by assuming that it formed in isolation and 
   then fell inside the Galactic potential well (e.g. Carraro \& Lia 2000; 
   Bekki \& Freeman 2003; Tsuchiya, Korchagin \& Dinescu 2004). The total mass 
   of the parent object ranges from 10$^8$~M$_\odot$ to some 10$^9$~M$_\odot$ 
   in those models. Alternative scenarios envisage an off-centre stellar 
   supercluster seed, which would have trapped older galactic field stars 
   during its formation process (Fellhauer, Kroupa \& Evans 2006). Other 
   possible explanations for the cluster origin -- namely, merging between two 
   or more smaller globulars or between a dwarf galaxy and a cluster (Icke \& 
   Alcaino 1988; Norris et al. 1997), or formation triggered by cloud-cloud 
   collisions (Tsujimoto \& Shigeyama 2003) -- have been put forward, but seem 
   less likely on the basis of the continuous trends of heavy elements-to-iron 
   and lanthanum-to-iron observed for S stars in $\omega$\,Cen (Vanture, 
   Wallerstein \& Suntzeff 2002, and references therein). Clear evidence 
   against the parent system evolving as a closed box has been put forward by 
   Ikuta \& Arimoto (2000).

   In this paper we deal with the chemical evolution of $\omega$\,Cen. We 
   discuss two possible scenarios of formation: in the first one, the 
   precursor of $\omega$\,Cen is a small system which evolves either as a 
   closed-box or with some exchange of matter with the surroundings; in the 
   second one, the cluster is the leftover of an ancient nucleated dwarf 
   galaxy swallowed by our Galaxy some 10 Gyr ago. We pay special attention to 
   the intriguing subject of helium enhancement in bMS stars. For the first 
   time, this critical issue is faced in the framework of a complete, 
   self-consistent chemical evolution model, taking all the relevant physics 
   into account. In Section~2 we list the observations that we use to 
   constrain our chemical evolution model. The model is presented in 
   Section~3. Section~4 summarizes the model results, that are discussed and 
   compared to previous investigations in Section~5. In Section~5 we also draw 
   our conclusions.

   \section{The nature of the chemical enrichment in Omega Centauri}

\begin{figure}
   \psfig{figure=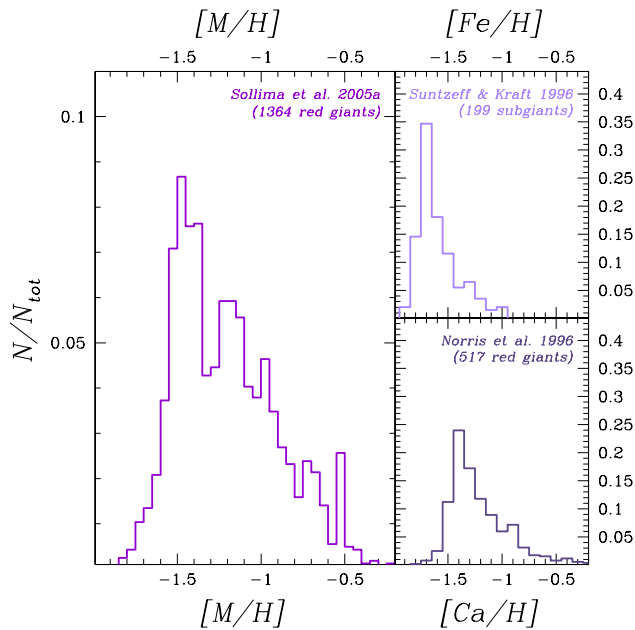,width=\columnwidth}
      \caption{ Shown are some of the observed metallicity distribution 
                functions for $\omega$\,Cen giants. General features of all 
		the observed distributions are the steep rise at low 
		metallicities and the high-metallicity tail (see text for 
		details).
              }
         \label{fig:mdfobs}
   \end{figure}

   It was recognized several years ago that simple models for cluster 
   enrichment can not reproduce the number of metal-rich stars observed in the 
   metallicity distribution function (MDF) of $\omega$\,Cen giants (Norris et 
   al. 1996; Suntzeff \& Kraft 1996; Ikuta \& Arimoto 2000). Based on Ca 
   abundances obtained from low-resolution spectra, Norris et al. (1996) found 
   a bimodal distribution with a metal-poor component, peaking at 
   [Ca/H]~$\simeq$ $-$1.4 dex and comprising nearly 80 per cent of the stars, 
   and a metal-rich one, comprising nearly 20 per cent of the stars at 
   [Ca/H]~$\simeq$ $-$0.9 dex (Fig.~\ref{fig:mdfobs}, lower right panel). 
   Suntzeff \& Kraft (1996) also found a sharp rise at low metallicities and a 
   high-metallicity tail, but no evidence for a secondary hump at higher 
   metallicities (Fig.~\ref{fig:mdfobs}, upper right panel).

%
%

   The sharp rise to a mean of [Fe/H]~$\simeq$~$-$1.6 dex and the long tail at 
   higher metallicities have been confirmed by a number of subsequent studies, 
   using both higher resolution spectroscopy and/or high quality photometry 
   (e.g. Hilker \& Richtler 2000; Frinchaboy et al. 2002; Sollima et al. 
   2005a; Stanford et al. 2006; Kayser et al. 2006). Some of these works also 
   show that the MDF is more complex than previously thought, with several 
   separate peaks identified in the observed distribution (e.g. Sollima et al. 
   2005a; Fig.~\ref{fig:mdfobs}, left panel). According to subgiant branch 
   (SGB) data, the different populations have ages comparable within 2 Gyr; 
   actually, they might be even coeval (Sollima et al. 2005b; see also Ferraro 
   et al. 2004). A wider age range, $\Delta t \simeq$ 2--4~Gyr, is inferred 
   from MSTO data (Stanford et al. 2006; see also Hilker et al. 2004 and 
   Kayser et al. 2006).

   The abundance ratios of different chemical species provide another 
   independent constraint on the evolutionary time-scales. As first discussed 
   by Lloyd Evans (1983), the enrichment of \emph{s}-process elements (e.g. 
   Ba, La) in the metal-rich $\omega$\,Cen stars is likely to be that of the 
   gas clouds out of which the stars formed. Smith et al. (2000) suggested 
   previous generations of low-mass asymptotic giant branch (AGB) stars as 
   polluters and argued in favour of a protracted period of star formation in 
   $\omega$\,Cen, of the order of 2--3 Gyr. Shorter time-scales are inferred 
   from the observed knee in the [$\alpha$/Fe] versus [Fe/H] relation, if Type 
   Ia supernovae (SNeIa) in $\omega$\,Cen restore the bulk of their iron to 
   the interstellar medium (ISM) on the same time-scale as in the solar 
   neighbourhood ($\sim$1~Gyr; Pancino et al. 2002). However, one must be 
   aware that the time-scale for the maximum enrichment by SNeIa in a specific 
   system depends strongly on the assumptions about the SN progenitors, 
   stellar lifetimes, initial mass function (IMF), star formation rate (SFR) 
   and, last but not least, possible metal-enriched gas outflows from the 
   system. It has been demonstrated (Matteucci \& Recchi 2001, and references 
   therein) that time-scales as long as $\sim$4--5 Gyr can be obtained with 
   suitable assumptions. We will come back to this issue later on, when 
   discussing model results in Section~\ref{sec:dis}.

   Gnedin et al. (2002) used simple dynamical modeling to demonstrate that if 
   $\omega$\,Cen had always evolved in isolation on its present orbit in the 
   Milky Way, it would have been unable to retain the \emph{s}-process-rich 
   wind material from AGB stars because of stripping from ram pressure during 
   many passages through the disc. Subsequent numerical simulations 
   demonstrated that an $\omega$\,Cen-like object can originate from a 
   nucleated dwarf galaxy intruding into the Milky Way. Both the orbital 
   parameters and the observed surface brightness profile of the present-day 
   $\omega$\,Cen are reproduced by a tidal disruption scenario where the 
   falling dwarf has its outer stellar envelope almost completely stripped, 
   whereas a central, dense nucleus still survives owing to its compactness 
   (Bekki \& Freeman 2003; Tsuchiya et al. 2004; Ideta \& Makino 2004). 
   Interestingly enough, also the number fraction (25 per cent) of the bMS 
   stars with extreme helium enhancement ($\Delta Y \approx$ 0.12--0.14; 
   Norris 2004; Piotto et al. 2005) can not be explained as long as they 
   originated from the ejecta of rMS stars initially within $\omega$\,Cen. 
   Rather, most of the helium-rich gas necessary to form the bMS had to come 
   from field stellar populations surrounding $\omega$\,Cen when it was the 
   compact nucleus of a Galactic dwarf satellite (Bekki \& Norris 2006). In 
   other words, the original total mass of the rMS population must have been 
   larger than the present-day one and most of the rMS stars must have been 
   removed from the proto-$\omega$\,Cen after their ejecta were used to form 
   the bMS population. Searches for tidal debris from $\omega$\,Cen's 
   hypothetical parent galaxy in the solar neighbourhood are possible in 
   principle (Dinescu 2002). Indeed, a distinct population of stars with 
   $\omega$\,Cen-like phase-space characteristics and metallicities consistent 
   with those of $\omega$\,Cen members emerges from catalogs of 
   metal-deficient stars in the vicinity of the Sun (Dinescu 2002; Meza et al. 
   2005).

   Though gas fueling from an ancient host galaxy stands as an attractive 
   hypothesis in many respects, to the best of our knowledge this possibility 
   has never been studied by means of fully self-consistent chemical evoution 
   models before.

   \section{The chemical evolution model}

   We follow the evolution of the abundances of several chemical species in 
   the gaseous medium out of which $\omega$\,Cen's stars form, in the case of 
   either a GC precursor or a dwarf spheroidal progenitor. The adopted 
   chemical evolution model is one zone, with instantaneous\footnote{Notice 
   that \emph{`instantaneous'} actually means \emph{`on time-scales shorter 
   than the adopted timestep for integration of the equations'.}} and complete 
   mixing of gas inside it and no instantaneous recycling approximation (i.e., 
   the stellar lifetimes are taken into account in detail). The GC precursor 
   has an initial mass of the same order of magnitude of the present one 
   (${\mathscr{M}}{\textrm{\scriptsize{$\omega$ Cen}}}$~= 2.5--5~$\times$ 
   10$^6$~M$_\odot$). Both a closed-box model and models where exchanges of 
   matter with the surroundings are allowed are analyzed. For the dwarf galaxy 
   precursor, when the computation starts the baryonic matter (cooling gas) is 
   embedded in a relatively massive 
   (${\mathscr{M}}_{\textrm{\scriptsize{dark}}}/
   {\mathscr{M}}_{\textrm{\scriptsize{bar}}}$~= 10), diffuse 
   ($R_{\textrm{\scriptsize{dark}}}/R_{\textrm{\scriptsize{eff}}}$~= 10), 
   virialized dark matter halo. Initially, the baryonic mass inside the dark 
   matter potential well is two orders of magnitude higher than the present 
   one. As soon as the star formation begins, the thermal energy of the gas 
   starts to increase as a consequence of multiple SN explosions, eventually 
   exceeding its binding energy\footnote{We compute the binding and thermal 
   energies of the gas according to the recipes of Bradamante, Matteucci \& 
   D'Ercole (1998). However, we adopt a typical efficiency of thermalization 
   from both Type II and Type Ia SNe of $\eta_{\textrm{\scriptsize{SNII}}}$ = 
   $\eta_{\textrm{\scriptsize{SNIa}}}$ = 0.20, rather than 0.03 (see Romano, 
   Tosi \& Matteucci 2006, and references therein).}. When this condition is 
   met, part of the gas escapes the galactic potential well; we assume that 
   this gas is definitively lost from the system.
   
   \subsection{Basic equations}

   We use the following basic equation (Tinsley 1980)
   \begin{equation}
     \frac{{\textrm d{\mathscr{G}}_i(t)}}{{\textrm d}t} = - X_i(t)\psi(t) + 
     {\mathscr{R}}_i(t) + \frac{{\textrm 
     d{\mathscr{G}}_i^{\textrm{\scriptsize{in}}}(t)}}{{\textrm d}t} - 
     \frac{{\textrm d{\mathscr{G}}_i^{\textrm{\scriptsize{out}}}(t)}}{{\textrm 
     d}t}
     \label{eq:base}
   \end{equation}
   to track the evolution of the fractional gas mass in the form of element 
   $i$ normalized to the initial gaseous mass, ${\mathscr{G}}_i(t) = X_i(t) 
   {\mathscr{M}}_{\textrm{\scriptsize{gas}}}(t)/
   {\mathscr{M}}_{\textrm{\scriptsize{bar}}}$. The quantity $X_i(t)$ 
   represents the abundance by mass of the element $i$ at the time $t$; by 
   definition, the summation over all the elements in the gas mixture is equal 
   to unity. The first term on the right hand side accounts for gas 
   consumption by star formation; the second term on the right hand side 
   refers to gas return by dying stars. All the complicated dependencies on 
   the adopted stellar initial mass function and lifetimes, SNIa progenitors 
   and stellar nucleosynthesis products hidden in the ${\mathscr{R}}_i(t)$ 
   term are not made explicit here; the interested reader can find all of them 
   discussed in considerable detail in Matteucci \& Greggio (1986). Suffice it 
   here to say that in this work we use an extrapolated Salpeter (1955) IMF or 
   a Scalo (1986) IMF, both normalized to unity over the 0.1--100~M$_\odot$ 
   stellar mass range, to show the predictions from both steep and flat, 
   although `standard', IMFs. The adopted stellar nucleosynthesis 
   prescriptions are discussed in Section~\ref{subsec:nuc}.

   In our models the SFR is a simple Schmidt's (1963) law:
   \begin{equation}
     \psi(t) = \nu {\mathscr{G}}^k(t).
   \end{equation}
   The quantity $\nu$ is the star formation efficiency, namely the inverse of 
   the typical time-scale for star formation, and is expressed in units of 
   Gyr$^{-1}$. The exponent $k$ is set to be 1.

   The last two terms in Equation~(\ref{eq:base}) account for any gas inflow 
   and/or outflow. For the closed system, all the gas available for star 
   formation is \emph{in situ} when the star formation begins at $t$~= 0, and 
   no infall from outside is considered. For the open systems, the rate of 
   gas infall is parametrized as
   \begin{equation}
     \frac{{\textrm d{\mathscr{G}}_i^{\textrm{\scriptsize{in}}}(t)}}{{\textrm 
     d}t}= \frac{X_{i}^{\textrm{\scriptsize{in}}} e^{-t/\tau}}{\tau (1 - 
     e^{-t_{\textrm{\scriptsize{now}}}/\tau})},
     \label{eq:inf}
   \end{equation}
   with $\tau$, the infall time-scale, set to be 0.5 Gyr, and 
   $X_{i}^{\textrm{\scriptsize{in}}}$, the abundances of the infalling gas, 
   set to their primordial values. In particular, we assume 
   $Y_{\textrm{\scriptsize{P}}}$~= 0.248, in agreement with the predictions of 
   the standard big bang nucleosynthesis (SBBN) theory and the constraints 
   from the cosmic microwave background (CMB; see Romano et al. 2003, and 
   references therein).

   The rate of gas loss via SN-driven large-scale outflows is different from 
   zero only for the open models, and simply proportional to the amount of gas 
   present at the time $t$:
   \begin{equation}
     \frac{{\textrm d{\mathscr{G}}_i^{\textrm{\scriptsize{out}}}(t)}}{{\textrm 
     d}t}= w_i X_{i}(t ) {\mathscr{G}}(t).
     \label{eq:out}
   \end{equation}
   The quantity $w_i$ is a free parameter which describes the efficiency of 
   the galactic wind; it is expressed in Gyr$^{-1}$ and may have different 
   values for different elements (e.g. Recchi, Matteucci \& D'Ercole 2001).

   \subsection{The evolutive context}

   According to Bekki \& Freeman (2003, and references therein), in the dwarf 
   progenitor scenario the initial stellar mass of $\omega$\,Cen's host was 
   significantly higher than that currently observed and can be estimated as
   \begin{equation}
     {\mathscr{M}}_{\textrm{\scriptsize{dwarf}}} = 
     \frac{{\mathscr{M}}_{\textrm{\scriptsize{$\omega$ Cen}}}}
	  {(1 - f_{\textrm{\scriptsize{lost}}})f_{\textrm{\scriptsize{n}}}},
   \end{equation}
   where $f_{\textrm{\scriptsize{n}}}$~= 0.05 is the mass fraction of the 
   compact nucleus and $f_{\textrm{\scriptsize{lost}}}$~= 0.2 is the stellar 
   mass fraction that gets lost through long-term ($\sim$10 Gyr) tidal 
   interaction with the Milky Way. If 
   ${\mathscr{M}}_{\textrm{\scriptsize{$\omega$ Cen}}} \simeq$ 
   5~$\times$~10$^6$~M$_\odot$ (Meylan et al. 1995), one gets 
   ${\mathscr{M}}_{\textrm{\scriptsize{dwarf}}}$~= 
   1.25~$\times$~10$^8$~M$_\odot$. Since  
   ${\mathscr{M}}_{\textrm{\scriptsize{$\omega$ Cen}}}$ 
   might be a factor of two lower (van de Ven et al. 2006), we run several 
   models, starting with an initial gaseous mass 
   ${\mathscr{M}}_{\textrm{\scriptsize{gas}}}(t = 0) = 
   {\mathscr{M}}_{\textrm{\scriptsize{bar}}}$~= 
   5~$\times$~10$^8$--10$^9$~M$_\odot$ and ending up with a stellar mass 
   ${\mathscr{M}}_{\textrm{\scriptsize{stars}}} (t$ = 3$) \simeq$ 
   5~$\times$~10$^7$--10$^8$~M$_\odot$. Given the similarity of the results, 
   in the case of the dwarf galaxy parent we will show only the predictions 
   for our most massive model ($\nu$~= 0.35 Gyr$^{-1}$, 
   $w_i^{\textrm{\scriptsize{max}}}$~$\simeq$ 5~Gyr$^{-1}$).

   Though in our models the star formation activity lasts $\sim$3 Gyr, it is 
   worth noting that most of the stars (nearly 75 per cent of the cluster 
   population) actually form during the first 1~Gyr. In the dwarf progenitor 
   scenario, while the star formation proceeds, the galaxy gets almost 
   completely depleted of its gas, partly owing to gas consumption by the 
   adopted long-lasting star formation activity, but most of all because of 
   efficient gas removal through the large-scale galactic outflows.

   The chemical properties of our $\omega$\,Cen progenitor systems are 
   discussed and compared with the available observations in 
   Section~\ref{sec:res}.
   
   \subsection{Nucleosynthesis prescriptions}
   \label{subsec:nuc}

   In this work we adopt the metallicity-dependent yields of van den Hoek \& 
   Groenewegen (1997) for single low- and intermediate-mass stars (LIMSs; 
   0.9~$\le$ $m$/M$_\odot$~$\le$ 8) and the yields of Nomoto et al. (1997) for 
   massive stars (13~$\le$ $m$/M$_\odot$~$\le$ 70). These latter are computed 
   for a solar chemical composition of the stars. The stellar yields are then 
   scaled to the current metallicity of the model by means of the production 
   matrix formalism (Talbot \& Arnett 1973). We (arbitrarily) use linear 
   interpolations and extrapolations to cover the 1--100 M$_\odot$ stellar 
   mass range. The effect of adopting different yield sets will be thoroughly 
   analyzed in a forthcoming paper (Romano et al., in preparation; notice that 
   changing the adopted yield sets is not expected to affect significantly 
   the main results presented in this paper).

   For stars evolving in binary systems which will give rise to SNIa 
   explosions, the nucleosynthesis prescriptions are from Iwamoto et al. 
   (1999). In our models, a substantial fraction of Cu and Zn is produced by 
   SNeIa, following the suggestions of Matteucci et al. (1993).

   \section{Results}
   \label{sec:res}

   \subsection{The closed-box picture}

   In this section we briefly discuss the results obtained in the framework of 
   the closed-box self-enrichment scenario, where any matter exchange between 
   the proto-$\omega$\,Cen and its environment is strictly forbidden.

%
   \begin{figure}
   \psfig{figure=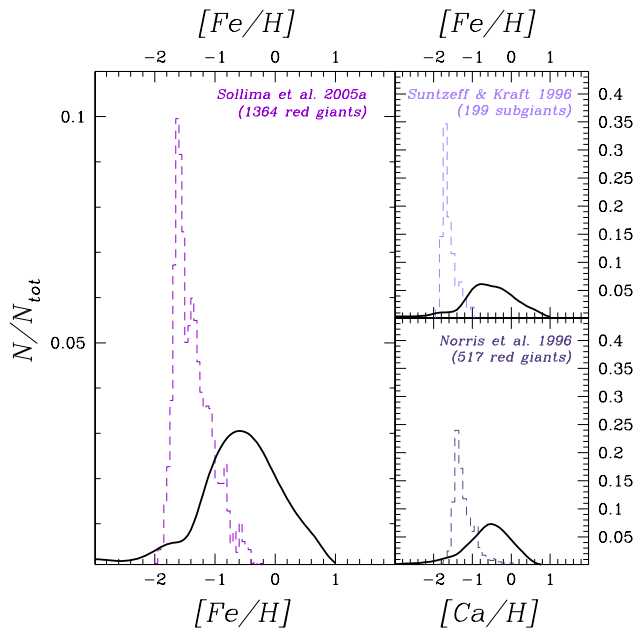,width=\columnwidth}
      \caption{ Comparison of observed (thin dashed histograms) and predicted 
               (thick solid lines) metallicity distribution functions of 
	       $\omega$\,Cen stars. Theoretical predictions refer to the 
	       closed-box self-enrichment scenario. The theoretical 
	       distributions are convolved with a Gaussian of dispersion 
	       $\sigma$~= 0.2 dex in order to (generously) take observational 
	       errors into account.
              }
         \label{fig:closed}
   \end{figure}
%

   In Fig.~\ref{fig:closed}, the theoretical metallicity distribution as a 
   function of [Fe/H] or [Ca/H] (thick solid lines) is compared with the 
   photometric and spectroscopic empirical ones (thin dashed histograms; 
   Norris et al. 1996; Suntzeff \& Kraft 1996; Sollima et al. 2005a). This 
   comparison shows the shortcomings of this simple picture. In the context of 
   a closed-box evolution, metallicities much higher than observed are quickly 
   attained, so that most of the stars form from matter with a chemical 
   composition from one tenth of solar to supersolar, at variance with the 
   observations.

   The results displayed in Fig.~\ref{fig:closed} are for the Scalo (1986) 
   IMF, that -- owing to its steeper slope for $m >$~2 M$_\odot$ with respect 
   to Salpeter's, $x$~= 1.7 rather than 1.35 -- allows a lower fraction of 
   high-mass stars to pollute the ISM with their metal-rich ejecta. Obviously, 
   by adopting a Salpeter IMF the predicted distribution goes towards even 
   higher metallicities. Flattening the IMF in the very low stellar mass 
   domain, as in a Kroupa et al.'s (1993) IMF, does not prevent the system 
   from suddenly reaching too much high metal abundances. Even in the 
   framework of extreme scenarios where the formation of SNIa progenitors is 
   suppressed and lower Fe yields from core collapse SNe are assumed (Chieffi 
   \& Limongi, in preparation), we still fail to reproduce the observed MDF. 
   Since within the scheme of a closed-box self-enrichment we can not fulfil 
   even such a basic observational constraint, we deem it meaningless to 
   analyze further the model results and switch to the `open' scenarios.

   Indeed, only if a substantial fraction of SN ejecta escapes from the 
   cluster, the observed MDF can be reproduced. Our result supports the 
   findings of Ikuta \& Arimoto (2000) that significant outflow from the 
   cluster is needed in order to reduce the effective yield per stellar 
   generation. Ikuta \& Arimoto (2000) obtained the best fit to the observed 
   MDF with a model involving gas outflow, infall at the very early stage of 
   chemical evolution and a bimodal IMF. However, the duration of star 
   formation for their best-fit model was only 0.28 Gyr, much shorter than 
   that inferred from current observations and assumed here.

   \subsection{A stripped dwarf galaxy?}

   An open, small-mass model hence offers a viable solution, but the 
   questionable assumption has to be made that, while SN products easily leave 
   the cluster, the stellar wind ejecta are completely retained and the 
   surrounding medium remains unperturbed. The current mass of $\omega$\,Cen 
   is unlikely to generate sufficient dynamical friction to modify its orbit 
   to its present small size (Majewski et al. 2000). Dynamical modeling, 
   however, points out that the long and complex star formation history of 
   $\omega$\,Cen is inconsistent with the cluster originating on its present 
   orbit: with a period of only 120 Myr (Dinescu et al. 1999), the frequent 
   disc crossings would have swept out all the intracluster gas very soon, 
   leading to a mono-metallicity system. Thus, the progenitor of $\omega$\,Cen 
   must have been a massive enough system to allow dynamical friction to drag 
   it to the inner Galactic regions (Bekki \& Freeman 2003). In light of these 
   considerations, in the following we discuss the results for a nucleated 
   dwarf hosting $\omega$\,Cen with initial mass 
   ${\mathscr{M}}_{\textrm{\scriptsize{bar}}}$~= 10$^9$~M$_\odot$. After a 
   3~Gyr evolution, the parent system has lost most of its gas through 
   galactic winds and ended up with a stellar mass 
   ${\mathscr{M}}_{\textrm{\scriptsize{stars}}}\sim$~10$^8$~M$_\odot$, which 
   is consistent with that of $\omega$\,Cen's parent galaxy according to the 
   computations of Bekki \& Freeman (2003).

   \subsubsection{Stellar metallicity distribution and age-metallicity 
                  relation}

%
   \begin{figure}
   \psfig{figure=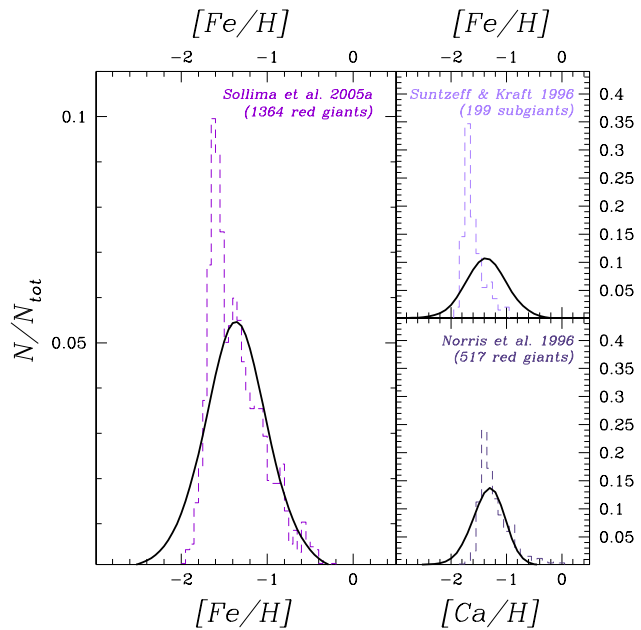,width=\columnwidth}
      \caption{ Comparison of observed (thin dashed histograms) and predicted 
               (thick solid lines) metallicity distribution functions of 
               $\omega$\,Cen stars. Theoretical predictions refer to the 
               evolutive picture where $\omega$\,Cen is the remnant of a 
	       larger system evolved in isolation and then accreted and 
	       partially disrupted by the Milky Way. The theoretical 
	       distributions are convolved with a Gaussian of dispersion 
	       $\sigma$~= 0.2 dex in order to (generously) take observational 
	       errors into account.
              }
         \label{fig:mdf}
   \end{figure}
%

   The observed MDF of long-lived stars in a galaxy is an important record of 
   its past evolution. In fact, the relative numbers of stars which formed at 
   any metallicity testify the interplay of fundamental processes such as star 
   formation, infall of gas from the surroundings and metal-enriched gas 
   outflows at any time. Reproducing the currently observed MDF of 
   $\omega$\,Cen's stars significantly restrains the free parameter space of 
   our dwarf galaxy model. We find that a fast early collapse coupled with an 
   intense (per unit mass) star formation activity, $\langle \psi 
   \rangle$~$\simeq$ 0.1~M$_\odot$ yr$^{-1}$ during the first 1~Gyr evolution, 
   gives rise to a distribution peaked at [Fe/H]~$\sim -$1.6, as observed 
   (Fig.~\ref{fig:mdf}, left panel), \emph{thanks to the strong galactic wind 
   which efficiently removes the metals from the proto-$\omega$\,Cen}. In our 
   model, the thermal energy of the gas exceeds its binding energy nearly 200 
   million years after the onset of the star formation, owing to multiple SN 
   explosions: the gas is then swept away by a strong galactic wind and the 
   star formation slowly fades. We impose that the wind is differential, 
   namely that the SN ejecta leave the galaxy more easily than the stellar 
   wind ejecta (see also Recchi et al. 2001). In particular, by assuming a 
   wind efficiency $w_i \simeq$ 5~Gyr$^{-1}$ for the SN ejecta, no stars with 
   [Fe/H]~$> -$0.4 dex are formed, in agreement with the observations 
   (Fig.~\ref{fig:mdf}, left panel). Such high outflow rates -- more than ten 
   times the SFR -- are often required to reproduce the high-quality data of 
   nearby dwarf spheroidals (Lanfranchi \& Matteucci 2003, 2004).  In 
   Fig.~\ref{fig:mdf} we compare the predictions of this model (thick solid 
   lines) with the same observed distributions of Figs.~\ref{fig:mdfobs} and 
   \ref{fig:closed}. Both the steep rise at low metallicities and the extended 
   metallicity tail are well reproduced; in particular, the agreement with the 
   up-to-date distribution of Sollima et al. (2005a) is strikingly good. This 
   is encouraging, but the model has to be tested against many more 
   observational constraints in order to prove its validity.

%
   \begin{figure}
   \psfig{figure=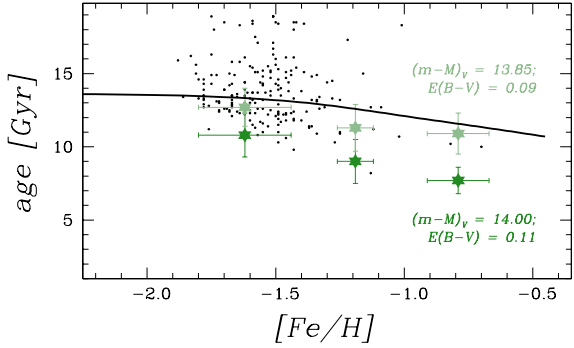,width=\columnwidth}
      \caption{ Predicted AMR for $\omega$\,Cen (thick solid line) compared to 
	        (i) the AMR inferred from a sample of $\sim$250 SGB stars 
	        with [Fe/H] errors less than 0.2 dex and age errors less than 
		2 Gyr, for two different choices of distance modulus and 
		reddening (stars; Hilker et al. 2004); (ii) the 
		age-metallicity diagram for the MSTO sample of Stanford et al. 
		(2006; dots, only stars with V $<$ 18 are considered). Notice, 
		however, that the most metal-rich stars at [Fe/H]~$> -$1.0 
		might have an accreted origin and ages comparable to that of 
		the main cluster population (see text for references).
              }
         \label{fig:amr}
   \end{figure}
%

   In a recent study by Hilker et al. (2004; see also Kayser et al. 2006), 
   newly derived spectroscopic abundances of iron for $\sim$400 $\omega$\,Cen 
   members have been used in combination with the location of the stars in the 
   CMD to infer the age-metallicity relation (AMR) of the system. The 
   suggested age spread is about 3~Gyr and there is some indication that the 
   AMR could level off above [Fe/H]~$\simeq -$1.0 dex. This is clearly seen in 
   Fig.~\ref{fig:amr}, where the stars with the error bars represent the AMR 
   for the subsample of stars with most reliable age and metallicity 
   determinations, for two choices of the reddening and distance modulus 
   values (see Hilker et al. 2004; their table~1). Evidence for the most 
   metal-rich stellar populations of $\omega$\,Cen being younger by 2--4~Gyr 
   than the most metal-poor one has been found also by Stanford et al. (2006) 
   from extensive Monte Carlo simulations on their MSTO data 
   (Fig.~\ref{fig:amr}, dots)\footnote{ Note that such an age difference seems 
   inconsistent with the results obtained by Ferraro et al. (2004) and Sollima 
   et al. (2005b) from the magnitude level, shape and extension of the turnoff 
   region.}. Although a large dispersion is present in the data, our model 
   clearly predicts the correct run of metallicity with age 
   (Fig.~\ref{fig:amr}, thick solid line). Notice that in our model \emph{the 
   SFR goes to zero after a 3~Gyr evolution, when [Fe/H]~$\simeq -$0.4 dex,} 
   because of the strong galactic outflows which efficiently remove any 
   residual gas from the galaxy (see previous paragraph).

   \subsubsection{Abundance ratios}
   \label{subsec:arat}

%
   \begin{figure}
   \hspace{.25cm}
   \psfig{figure=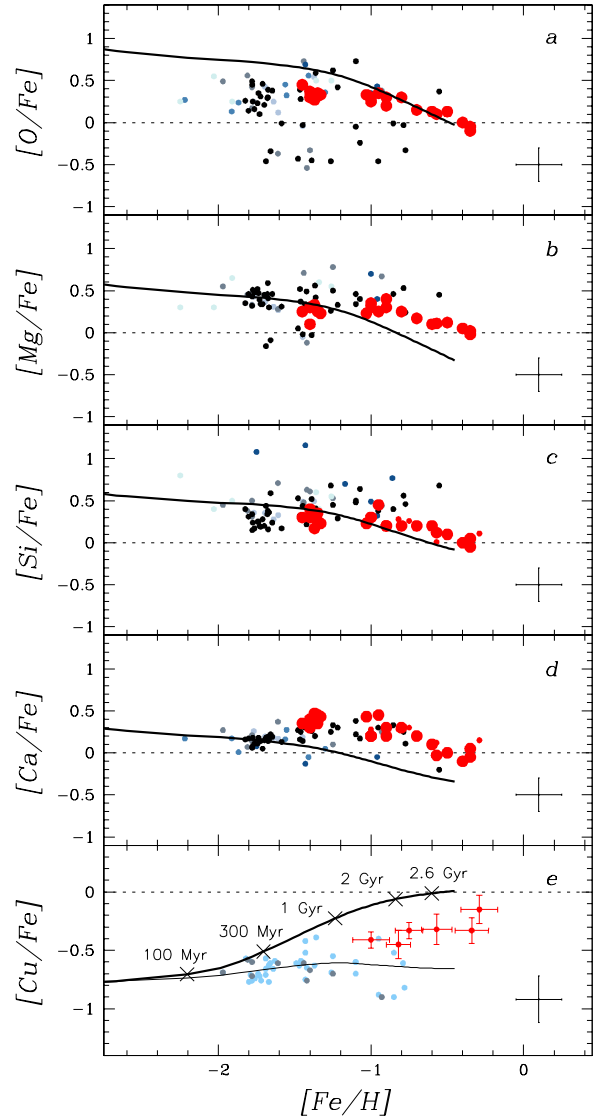,width=\columnwidth}
      \caption{ Predicted (solid lines) versus observed (filled circles; see 
	        text for references) abundance ratios of several chemical 
		species to iron as a function of [Fe/H]. The $\alpha$ elements 
		oxygen (panel a), magnesium (panel b), silicon (panel c) and 
		calcium (panel d) are shown, as well as the iron-peak element 
		copper (panel e). Conservative errors are shown in the right 
		hand corner of each panel. The crosses in panel e mark the 
		time elapsed since the beginning of star formation (0.1, 0.3, 
		1, 2, 2.6 Gyr); they are not superimposed on the track in all 
		panels in order to avoid confusion and overcrowded plots.
              }
         \label{fig:abun}
   \end{figure}
%

   In Fig.~\ref{fig:abun} we display our predictions for several abundance 
   ratios as a function of [Fe/H] (thick solid lines). The data displayed in 
   Fig.~\ref{fig:abun} (circles) have been collected from the literature. Data 
   from high-resolution optical spectra (Fran\c cois, Spite \& Spite 1988; 
   Brown \& Wallerstein 1993; Norris \& Da Costa 1995; Smith et al. 1995, 
   2000; Cunha et al. 2002; Pancino et al. 2002; Vanture et al. 2002) are 
   shown as small circles. Data from low- and medium-resolution infrared 
   spectra (Origlia et al. 2003) are shown as big circles.

   We have checked that all literature abundance ratios are in reasonable 
   agreement with each other (we consider two studies in reasonable agreement 
   when the abundances of the stars in common do not differ, on average, by 
   more than 0.1~dex, which is the typical uncertainty of the measurements). 
   As far as [Fe/H] is concerned, most papers agree with each other with the 
   exceptions of Fran\c cois et al. (1988), Pancino et al. (2002), Vanture et 
   al. (2002) and Origlia et al. (2003), all having [Fe/H]~$\sim$ 0.2~dex 
   lower, which is a marginal ($\sim$~2\,$\sigma$) discrepancy. We therefore 
   compared atmospheric parameters ($T_{\textrm{\scriptsize{eff}}}$, $\log 
   \mathit{g}$ and $v_t$), atomic data ($\log \mathit{gf}$) and equivalent 
   width (EW) measurements among the above studies, to find possible 
   explanations for the discrepancies. In the case of Fran\c cois et al., the 
   turbolent velocities are significantly lower ($\sim$ 0.5~km s$^{-1}$) and 
   the adopted solar composition is significantly higher, 
   $\log \varepsilon$(Fe)$_\odot$~= 7.67. Vanture et al. have instead a much 
   lower $v_t$ (by 0.4~km s$^{-1}$) than average, and $\log \mathit{gf}$ 
   slightly lower ($\sim$~0.06~dex). In the case of Pancino et al., all 
   parameters appear in good agreement with the other literature sources, but 
   for the only star in common (ROA\,371) there is an average difference in EW 
   of $\sim$~6.5~m\AA. All these factors together suggest that we should 
   revise the Fran\c cois et al., Pancino et al. and Vanture et al. [Fe/H] 
   values upwards by 0.2~dex. Origlia et al. have only stars in common with 
   Pancino et al., with differences in $T_{\textrm{\scriptsize{eff}}}$ 
   ($\sim$~150~K),$\log \mathit{g}$ ($\sim$~0.3~dex) and $v_t$ ($\sim$~0.3~km 
   s$^{-1}$), but very similar abundances. Even if we do not have the tools to 
   fully understand the cause of the discrepancies (the method, resolution and 
   spectral range employed by Origlia et al. are very different from the rest 
   of the papers), we choose to revise the [Fe/H] values upwards by 0.2~dex, 
   in conformity with what done for Fran\c cois et al. (1988), Pancino et al. 
   (2002) and Vanture et al. (2002).

   Concerning the $\alpha$-elements, there is a good agreement among various
   studies, except for a few notable exceptions. Oxygen is a very difficult 
   element to measure, since only one line, [O\,{\small I}] at 6300~\AA, is 
   used by most authors, that lies in a region plagued by atmospheric 
   O$_{\textrm{{\scriptsize{2}}}}$ absorption and is blended with a 
   Ni\,{\small I} and a Sc\,{\small II} line. However, some authors also 
   include the weak 6363~\AA\ line in their analysis, while only part of the 
   authors perform a full spectral synthesis on the region. In spite of this, 
   the abundance determinations are in reasonable agreement with each other, 
   with the marginal exception of Vanture et al. (2002), that we choose to 
   leave as it is, since the overall scatter in [O/Fe] is significantly larger 
   than for other elements (see Fig.~\ref{fig:abun}). For Mg\,{\small I}, 
   Smith et al. (2000) have an abundance $\sim$~0.2~dex higher than average, 
   which is partly explained by the lower $\log \mathit{gf}$ ($\sim$~0.1~dex) 
   of the employed lines. We therefore lower the Smith et al. [Mg/Fe] 
   abundances by 0.1~dex, bringing them in reasonable agreement with other 
   determinations. Silicon abundances are a tricky business, since there is a 
   general disagreement among authors (only Fran\c cois et al. 1988 and Smith 
   et al. 2000 appear to be on the same scale), but no clear cause for the 
   discrepancies emerges from our analysis, so we choose to apply no 
   correction for this particular element. The Ca\,{\small I} determination by 
   Norris \& Da Costa (1995) is $\sim$~0.2~dex higher than all other studies 
   considered here, that is fully taken into account by the $\sim$~0.2~dex 
   lower $\log \mathit{gf}$ values adopted by those authors, so we revise 
   Norris \& Da Costa's [Ca/Fe] downwards by $\sim$~0.2~dex.

   Finally, only two papers have studied Cu in $\omega$\,Cen so far, namely 
   Pancino et al. (2002) and Cunha et al. (2002), having only one star in 
   common, ROA\,371. The linelist for the synthesis of the 
   Cu\,{\small I}~5782~\AA\ line used is different in the two papers. The 
   adopted atmospheric parameters are very similar, but the solar copper 
   abundances differ by 0.09~dex. All this together makes an $\sim$~0.2~dex 
   difference between the two studies of ROA\,371 ([Cu/Fe]~= $-$0.33 in 
   Pancino et al. and [Cu/Fe]~= $-$0.52 in Cunha et al.) as can be seen from 
   fig.~6 of Cunha et al. (2002). However, if we adopt the Grevesse \& Sauval 
   (1998) solar composition for Cunha et al., as in Fig.~\ref{fig:abun}, the 
   discrepancy rises to 0.3~dex ([Cu/Fe]~= $-$0.61 in Cunha et al.). We choose 
   to use the same solar abundance (Grevesse \& Sauval 1998), to put into 
   evidence the real discrepancy between the two studies, which reflects the 
   actual uncertainty in the derivation of Cu abundances.

   Theoretical values are normalized to the solar ones of Grevesse \& Sauval 
   (1998). Elements such as carbon and nitrogen are not considered, as their 
   original abundances could be easily altered in the atmospheres of the 
   sampled giant stars\footnote{ Actually, a large scatter is present also in 
   the abundances of oxygen. Although commonly explained as the signature of 
   processing in the CNO cycle (e.g. Norris \& Da Costa 1995), it is worth 
   stressing that it could be at least partly due to the wealth of 
   observational problems discussed above.}. The neutron capture elements Y, 
   Ba, La and Eu deserve special attention and will be treated in detail 
   elsewhere (Romano et al., in preparation).

   It can be seen that the model reproduces satisfactorily the decreasing 
   trend of various [$\alpha$/Fe] ratios with time (metallicity), with the 
   possible exception of Mg, which is underestimated for [Fe/H]~$> -$1.0. 
   Appropriate adjustment of the adopted nucleosynthesis prescriptions and IMF 
   slope would make the model predictions fit the data even better. Indeed, 
   Fran\c cois et al. (2004) find that the Nomoto et al. (1997) yields of Mg 
   we are using need significant corrections to best fit the abundance data of 
   very metal-poor Galactic halo stars, and the stellar mass function is known 
   to flatten in the very-low stellar mass domain (e.g. Kroupa et al. 1993; 
   Chabrier 2003). However, what really matters here is to provide an overall, 
   coherent interpretative framework for the whole body of data for 
   $\omega$\,Cen, rather than to attempt to precisely fit a specific 
   observational constraint.

   The high ratios of [O/Fe], [Mg/Fe], [Si/Fe] and [Ca/Fe] for [Fe/H]~$< -$1.2 
   clearly point to SNeII as the major drivers of the chemical enrichment. 
   Yet, the indication of a decrease of these abundance ratios in the 
   intermediate and metal-rich subpopulations (Pancino et al. 2002; Origlia et 
   al. 2003) are the unmistakable sign of significant Type Ia SN pollution, 
   occuring at later times but before star formation stops.

%
   \begin{figure}
   \psfig{figure=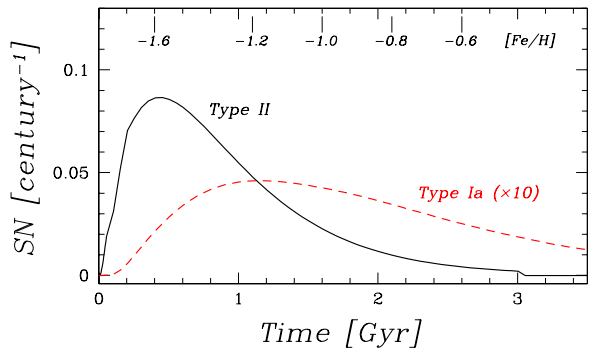,width=\columnwidth}
      \caption{ Theoretical Type II (solid line) and Type Ia (dashed line) SN 
                rates (number per century) for the $\omega$\,Cen progenitor. 
		The SNIa rate is multiplied by a factor of ten to make it 
		clearly visible.
              }
         \label{fig:sn}
   \end{figure}
%

   As far as copper is concerned, it is worth stressing that the flat 
   behaviour of the [Cu/Fe] data indicates no evolution in the copper-to-iron 
   yields over much of the chemical evolution within $\omega$\,Cen (see 
   Fig.~\ref{fig:abun}, panel e, circles, and Cunha et al. 2002, their 
   fig.~6). Yet, from Fig.~\ref{fig:abun}, panel e, it is seen that our 
   chemical evolution model (thick solid curve) overpredicts the 
   copper-to-iron ratio over the whole metallicity range. The predicted 
   behaviour of [Cu/Fe] results from the assumption that the major 
   astrophysical site for the synthesis of Cu are SNeIa (Matteucci et al. 
   1993), and these stellar factories happen to contribute to the chemical 
   enrichment of $\omega$\,Cen already at the lowest metallicities. This is 
   apparent from Fig.~\ref{fig:sn}, where we show the Type II (solid line) and 
   Type Ia (dashed line) SN rates predicted by our model for $\omega$\,Cen. 
   The SNIa rate has been multiplied by a factor of ten to make it clearly 
   visible. SNeII appear soon (a few million years) after their massive 
   progenitors are born, thus their rate closely follows the SFR. SNeIa, 
   instead, come from intermediate- to low-mass stellar progenitors in binary 
   systems (Matteucci \& Greggio 1986), which explode on varying time-scales: 
   they start to contribute significantly to the chemical enrichment of the 
   proto-$\omega$\,Cen $\sim$400 Myr after the beginning of star formation, 
   when the ISM has attained a metallicity of [Fe/H]~$\simeq -$1.6 
   (Fig.~\ref{fig:sn}, upper x axis), while the maximum enrichment takes place 
   only several Myr later, at $t \simeq$ 1.1 Gyr, when [Fe/H]~$\simeq 
   -$1.2\footnote{Notice that a bimodal distribution of the delay times for 
   the explosion, with \emph{`prompt'} and \emph{`tardy'} events, has recently 
   proven to best match present SNIa data (Mannucci, Della Valle \& Panagia 
   2006).}. Later on, at $t \simeq$ 2.6 Gyr ([Fe/H]~$\simeq -$0.6), the 
   contribution from SNeIa nearly equals that from SNeII, but only a few stars 
   form from this SNIa-enriched material before the star formation stops. It 
   is worth emphasizing here that the metal-enriched outflow lengthens the 
   enrichment time-scale, i.e. the time that it takes for the ISM to reach a 
   given metallicity: in the absence of such an enriched wind, it would take 
   only 0.7 Gyr for the ISM to reach a metallicity of [Fe/H]~$\simeq -$1.2 and 
   1.8 Gyr to reach [Fe/H]~$\simeq -$0.6.

   In their 1993 paper, in order to fit the solar neighbourhood data, 
   Matteucci and coworkers adopted Cu yields from SNeIa a factor of 100 higher 
   than predicted by current SNIa models. When adopting the lower Cu yields 
   from SNeIa predicted by Iwamoto et al. (1999), we find a flat behaviour of 
   [Cu/Fe] versus [Fe/H] in $\omega$\,Cen (Fig.~\ref{fig:abun}, panel e, thin 
   solid curve). While this is consistent with the data of Cunha et al. 
   (2002), it does not account for the rise for [Fe/H]~$> -$1.0 pointed out by 
   Pancino et al. (2002). Clearly, the issue of copper production in stars 
   need to be further investigated. Apart from the flatness, an even more 
   striking feature is the overall Cu deficiency in conjunction with a strong 
   \emph{s-}process enhancement. The same chemical signature appears to 
   characterize the Sagittarius dwarf spheroidal galaxy, thus lending support 
   to the idea that $\omega$\,Cen is the remaining nucleus of an accreted 
   dwarf Galactic satellite (McWilliam \& Smecker-Hane 2005).

   \subsection{Helium enrichment in $\bmath{\omega}$\,Cen}
   \label{subsec:reshe}

   In this section, we explore the issue of the helium enrichment in 
   $\omega$\,Cen within the picture where this cluster is the compact survivor 
   of a larger satellite system ingested by the Milky Way many Gyr ago.

%
   \begin{figure}
   \psfig{figure=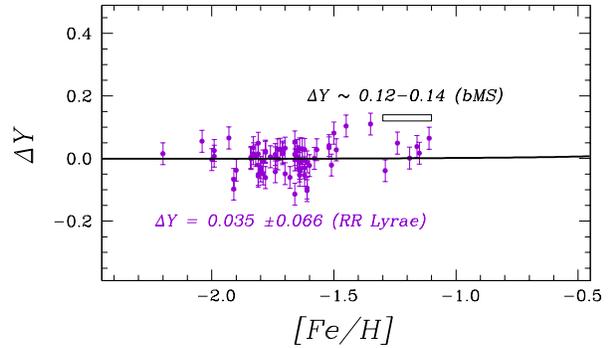,width=\columnwidth}
      \caption{ Relative He abundance as a function of metallicity. The 
                theoretical relation (thick line) is compared to the empirical 
		one obtained from RR Lyr\ae \ star data (filled circles; see 
		Sollima et al. 2006a, and references therein, for details 
		about the derivation of the relative He abundances in these 
		variable stars via the mass-luminosity parameter $A$). The box 
		represents the level of He enhancement required in order to 
		explain the bMS data (Norris 2004; Piotto et al. 2005).
              }
         \label{fig:dy}
   \end{figure}
%

   In Fig.~\ref{fig:dy} we compare the theoretical $\Delta Y$ versus [Fe/H] 
   relation (thick solid line) to the data (box and filled circles). When 
   using standard stellar yields (see Section~\ref{subsec:nuc}), after a 3~Gyr 
   evolution we obtain a negligible increase of the He abundance in the ISM 
   ($\Delta Y$~= 0.01), thus underestimating by an order of magnitude the 
   level of He enhancement required to explain the bMS and horizontal branch 
   (HB) data ($\Delta Y$~= 0.12--0.14; Norris 2004; Piotto et al. 2005). Yet, 
   our model predictions are fully consistent with the tiny change in helium 
   abundance ($\Delta Y$~= 0.035~$\pm$~0.066) over a wide metallicity range 
   ($-$2.2~$\le$ [Fe/H]~$\le -$1.1) implied by the observations of 74 RR 
   Lyr\ae \ variables by Sollima et al. (2006a). The fact that there is a 
   population of metal-intermediate RR Lyr\ae \ stars in $\omega$\,Cen with 
   luminosity and pulsational properties that are incompatible with a 
   significant helium overabundance adds new complexity to the peculiar 
   chemical features of $\omega$\,Cen, as it entails the coexistence of two 
   populations with similar metallicities but very different helium abundances 
   in the cluster (Sollima et al. 2006a).

\begin{figure}
   \psfig{figure=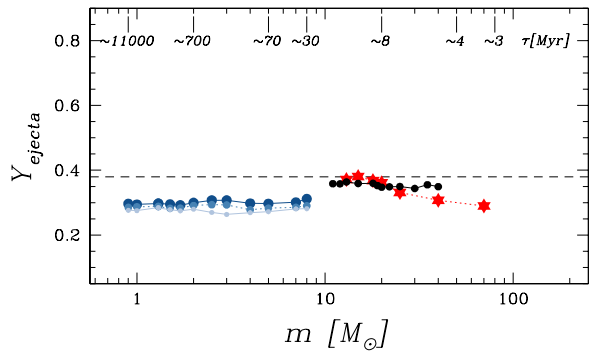,width=\columnwidth}
      \caption{ Mass fraction of He in the ejecta of LIMSs as well as massive 
	       stars as a function of stellar mass. 
	       $Y_{\textrm{\scriptsize{ejecta}}}$ values were calculated using 
	       the tables by van den Hoek \& Groenewegen (1997) for LIMSs 
	       and Woosley \& Weaver (1995; filled circles) and Nomoto et al. 
	       (1997; stars) for massive stars. In the case of van den Hoek \& 
	       Groenewegen's yields, the bigger the size of the symbol, the 
	       higher the initial metallicity of the star ($Z$~= 0.001, 0.004, 
	       0.008). For comparison, the $Y$ value suggested for the bMS 
	       stars of $\omega$\,Cen is also displayed as a dashed line. The 
	       adopted stellar lifetimes are reported (for a few objects) on 
	       the upper x axis.
              }
         \label{fig:yields}
   \end{figure}

\begin{figure}
   \psfig{figure=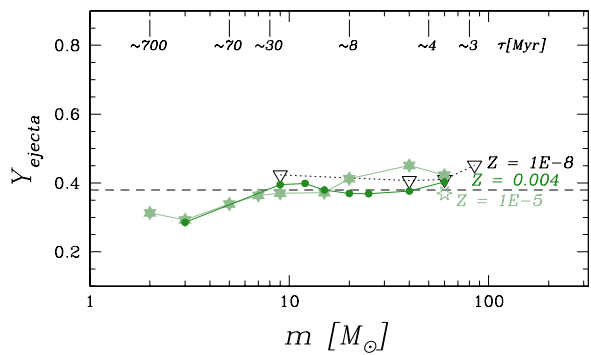,width=\columnwidth}
      \caption{ Mass fraction of He in the ejecta of rotating stars as a 
	       function of stellar mass. $Y_{\textrm{\scriptsize{ejecta}}}$ 
	       values were calculated using the tables by Meynet \& Maeder 
	       (2002), Hirschi (2006) and Meynet et al. (2006) for different 
	       initial metallicities and/or rotational velocities of the 
	       stars. Upside-down triangles: $Z$~= 10$^{-8}$; stars: $Z$~= 
	       10$^{-5}$; circles: $Z$~= 0.004. Empty symbols: 
	       $v_{\textrm{\scriptsize{ini}}}$~= 800~km s$^{-1}$; filled 
	       symbols: $v_{\textrm{\scriptsize{ini}}}$~= 300~km s$^{-1}$, 
	       except for the 9 M$_\odot$ and 40 M$_\odot$ stars at $Z$~= 
	       10$^{-8}$, for which $v_{\textrm{\scriptsize{ini}}}$~= 500~km 
	       s$^{-1}$ and 700~km s$^{-1}$, respectively. For comparison, the 
	       $Y$ value suggested for the bMS stars of $\omega$\,Cen is also 
	       displayed as a dashed line. The adopted stellar lifetimes are 
	       reported (for a few objects) on the upper x axis.
              }
         \label{fig:yrot}
   \end{figure}

   The stars on the blue side of the MS show no spread in metallicity (Piotto 
   et al. 2005), thus suggesting that they formed from a well-homogenized 
   medium. On the other hand, in order to elevate $Y$ from its primordial 
   value 0.248 to $\sim$0.40 one must assume that the material from which the 
   bMS stars formed was made up almost entirely of pure ejecta from previous 
   stellar generations (Norris 2004). In fact, as long as the fresh ejecta of 
   dying stars are diluted and mixed up with pre-existing gas in the framework 
   of a homogeneous chemical evolution model, there is no way of attaining a 
   significant helium enrichment of the ISM by the time the metallicity 
   increases from [Fe/H]~$\sim -$2.0 to $\sim -$1.2 dex (see 
   Fig.~\ref{fig:dy}, solid line). This result is independent of the adopted 
   stellar yields, as can be immediately understood from an inspection of 
   Figs.~\ref{fig:yields} and \ref{fig:yrot}. There, we show the mass fraction 
   of He in the ejecta of LIMSs and massive stars as a function of stellar 
   mass, calculated as
   \begin{equation}
     Y_{\textrm{\scriptsize{ejecta}}} 
	       = Y_{\textrm{\scriptsize{ini}}} + 
	         \frac{mp_{\textrm{\scriptsize{He}}}}{m - 
	         m_{\textrm{\scriptsize{rem}}}},
   \end{equation}
   where $Y_{\textrm{\scriptsize{ini}}}$ is the initial helium abundance of 
   the star, $p_{\textrm{\scriptsize{He}}}$ is the stellar yield, namely the 
   fractional mass of the star of initial mass $m$ which is restored to the 
   ISM in the form of newly produced He, and $m_{\textrm{\scriptsize{rem}}}$ 
   is the mass of the remnant.
%
%
%
%
   The quantities displayed in 
   Fig.~\ref{fig:yields} have been computed using the tables of van den Hoek 
   \& Groenewegen (1997) for LIMSs and Nomoto et al. (1997; stars) and Woosley 
   \& Weaver (1995; filled circles) for massive stars. In the case of van den 
   Hoek \& Groenewegen's yields, the bigger the size of the symbol, the higher 
   the initial metallicity of the star ($Z$~= 0.001, 0.004, 0.008). The 
   quantities displayed in Fig.~\ref{fig:yrot} have been computed using the 
   tables of Meynet \& Maeder (2002), Hirschi (2006) and Meynet et al. (2006) 
   for rotating stars, for different initial metallicities and/or rotational 
   velocities of the stars (see figure caption). We refer to those authors for 
   details about the stellar model assumptions. In the case of the standard 
   yields (Fig.~\ref{fig:yields}), $Y_{\textrm{\scriptsize{ejecta}}}$ is 
   always below the quantity needed to explain the bMS data. Rotating stars, 
   instead, do actually (slightly) exceed $Y_{\textrm{\scriptsize{ejecta}}}$~= 
   0.38 in the high-mass domain, but below 8~M$_\odot$ they always contribute 
   a much lower He amount (Fig.~\ref{fig:yrot}). Since, according to the 
   Salpeter IMF we are using, in each stellar generation about 85 per cent of 
   the mass falls in the 0.1--8~M$_\odot$ mass range, it is easy to guess that 
   after the $\sim$1~Gyr evolution needed to achieve [Fe/H]~$\simeq -$1.2 in 
   the ISM, the average helium abundance of the medium out of which the bMS 
   stars form will be definitely lower than required. Indeed, when including 
   the yields of the Geneva group in our homogeneous chemical evolution model, 
   we always find $Y$~$<$ 0.30 ($\Delta Y$~$<$ 0.05). We will come back to the 
   issue of the helium enrichment in $\omega$\,Cen in a future paper (Romano 
   et al., in preparation).

   \section{Discussion and conclusions}
   \label{sec:dis}

   In this work we study the formation and evolution of the anomalous globular 
   cluster $\omega$\,Cen. The constraints established by the wealth of very 
   good-quality abundance data for its member stars collected over the last 
   decade significantly restrain the evolutive picture. We show that in the 
   framework of the closed-box self-enrichment scenario (still an often used 
   approximation) the metallicity distribution function of $\omega$\,Cen's 
   stars can not be reproduced. On the other hand, the main chemical 
   properties of $\omega$\,Cen are nicely reproduced if it is the compact 
   remnant of a dwarf spheroidal galaxy evolved in isolation and then accreted 
   -- and partly disrupted -- by the Milky Way. This evolutive picture has 
   already allowed different authors to explain the present-day peculiar 
   location near the Galactic centre and surface brightness profile of 
   $\omega$\,Cen (see, e.g., Bekki \& Freeman 2003; Tsuchiya et al. 2004; 
   Ideta \& Makino 2004); our analysis gives further strength to the accreted 
   dwarf picture by proving that the ingested satellite would also have 
   similar chemical properties to $\omega$\,Cen. Indeed, by assuming a 
   relatively long-lasting star formation activity (though with most of the 
   stars forming within 1 Gyr), standard IMF and standard stellar yields, our 
   models satisfactorily reproduce several observed abundance ratios as a 
   function of [Fe/H].

   Regarding the helium abundances of $\omega$\,Cen, a great deal of work has 
   been recently devoted to the enigma of the anomalous helium-to-metal 
   enrichment of bMS and hot HB stars (Norris 2004; Lee et al. 2005; Piotto et 
   al. 2005; Bekki \& Norris 2006; Maeder \& Meynet 2006). Stars which 
   populate the `blue hook' region could have suffered an unusually large mass 
   loss on the RGB and, consequently, experienced the helium core flash while 
   descending the white dwarf cooling curve (Moehler et al. 2002). Therefore, 
   it is not clear which fraction of the He enhancement truly reflects the 
   pristine abundance. Stars on the bMS, instead, likely give more trustworthy 
   indication: they point to $\Delta Y$~= 0.12--0.14 ($\Delta Y/\Delta 
   Z$~$\ge$ 70). Recent observations of RR Lyr\ae \ stars suggest that a 
   population with a normal (i.e. nearly primordial) helium content inhabits 
   the cluster as well, thus complicating the overall picture (Sollima et al. 
   2006a). Our homogeneous chemical evolution model, when adopting a 3~Gyr 
   long star formation history, standard yields and Salpeter's IMF, predicts 
   that the helium abundance in the ISM almost does not change during the 
   cluster's progenitor evolution, in agreement with the RR Lyr\ae \ star 
   data, but in sharp disagreement with the abundances inferred for the bMS. 
   We compare existing helium yields for low-metallicity rotating stars to the 
   outputs of standard stellar evolution which does not take rotation into 
   account, and find that the winds of massive stars of intermediate rotation 
   velocities do indeed eject significant amounts of He (see also Maeder \& 
   Meynet 2006). However, once weighted with a normal IMF, the helium yield of 
   a stellar generation turns out to be lower than that required to explain 
   the bMS anomalies (see Section~\ref{subsec:reshe}). It is worth noticing at 
   this point that even if the escape of SN ejecta through galactic winds 
   helps to keep the global metal content of the galaxy low, thus favouring 
   high $\Delta Y/\Delta Z$ values, it also leads to longer enrichment 
   time-scales (see Section~\ref{subsec:arat}), thus allowing low- and 
   intermediate-mass stars to spread their relatively helium-poor gas through 
   the ISM. Ad hoc scenarios in which the winds of massive stars remain 
   confined and pollute only the central regions where the He-rich MS stars 
   are actually more concentrated may represent a way out of the problem. 
   Helium diffusion has recently been proposed as a viable mechanism to 
   significantly increase the helium content of protostellar clouds (Chuzhoy 
   2006).

   Assuming an IMF strongly biased towards massive stars could also solve the 
   problem, but current observational evidence seems to favour a Salpeter-like 
   IMF in dwarf spheroidals as well as Galactic globulars (Wyse 2005, and 
   references therein). Another option would be that the bMS was formed apart 
   as a star cluster and then merged with the rMS in the central region of 
   $\omega$\,Cen's host galaxy, a possibility first pointed out by Bekki \& 
   Norris (2006). We note that if the star cluster originated mostly from 
   matter ejected by slow winds from rotating massive stars with [Fe/H]~$\sim 
   -$1.2, then it would have both the high helium abundance and uniform 
   metallicity inferred from bMS observations. Measuring the helium abundances 
   of stars more metal-rich than [Fe/H]~$\sim -$1.0 would help us to 
   discriminate between a scenario where the helium abundance in the ISM 
   monotonically increases with time and a scenario where the formation of a 
   helium-rich population is a fortuitous event occurred on a short time-scale 
   under very special conditions. We will examine these scenarios in detail in 
   a forthcoming paper, together with the most updated (even ad hoc) 
   nucleosynthesis prescriptions.

   \section*{Acknowledgments}

   We thank L. Stanford for kindly providing her data in machine-readable 
   format.

\bsp

\label{lastpage}

\end{document}